\documentclass[aps,prd,showpacs,amsmath,amssymb,preprintnumbers]{revtex4}
\usepackage{epsfig}
\usepackage{amsmath}
\newcommand{\vect}{\left ( \begin{array}{c}}
\newcommand{\evect}{\end{array} \right )}
\usepackage{epsfig}
\usepackage{psfrag}
\usepackage{graphicx}
\usepackage{dcolumn}
\usepackage{bm}
\def\fsl#1{\setbox0=\hbox{$#1$}                 
   \dimen0=\wd0                                 
   \setbox1=\hbox{/} \dimen1=\wd1               
   \ifdim\dimen0>\dimen1                        
      \rlap{\hbox to \dimen0{\hfil/\hfil}}      
      #1                                        
   \else                                        
      \rlap{\hbox to \dimen1{\hfil$#1$\hfil}}   
      /                                         
   \fi}                                         %

\begin{document}

\title{On Melting Temperature of Heavy Quarkonium  with the AdS/CFT Implied  Potential}

\begin{abstract}
The quarkonium states in a quark-gluon plasma is examined with the
potential implied by AdS/CFT duality. Both the vanila
AdS-Schwarzschild metric and the one with an infrared cutoff are
considered. The calculated dissociation temperatures for $J/\psi$
and $\Upsilon$ are found to agree with the lattice results within a
factor of two.

\end{abstract}

\author{Defu Hou}
\email[E-mail:~] {hdf@iopp.ccnu.edu.cn} \affiliation{Institute of
Particle Physics, Huazhong Normal University, Wuhan 430079, China}
\author{Hai-cang Ren}
\email[E-mail:~] {ren@summit.rockefeller.edu} \affiliation{Physics
Department, The Rockefeller University, 1230 York Avenue, New York,
NY 10021-6399} \affiliation{Institute of Particle Physics, Huazhong
Normal University, Wuhan 430079, China}



\maketitle


The heavy quarkonium dissociation is one of the important signals of
the formation of QGP in RHIC. The subject has been explored
extensively on a lattice \cite{satz,review,karsch}. In the
deconfinement phase of QCD, the range of the binding potential
between a quark and an antiquark is limited by the screening length
in a hot and dense medium, which decreases with an increasing
temperature. Beyond the dissociate temperature, $T_d$, the range of
the potential is too short to hold a bound state and the heavy
quarkonium will melt. The lattice simulation of the quark-antiquark
potential and the spectral density of hadronic correlators  yield
consistent picture of the quarkonium dissociation and the numerical
values $T_d$.  On the other hand,AdS/CFT duality provides a new
avenue towards a qualitative or even semi-quantitative understanding
of the non-perturbative aspect of a quantum field theory
\cite{maldacena1, witten}. It is conjectured that a string theory
formulated in $AdS_5\times S_5$ is dual to the $N=4$ supersymmetric
Yang-Mills theory (SUSY YM) on the boundary. In particular, the low
energy limit of the classical string theory, the supergravity in
$AdS_5\times S_5$ corresponds to the supersymmetric Yang-Mills
theory at large $N_c$ and large 't Hooft coupling $ \lambda\equiv
N_cg_{\rm YM}^2 $. Remarkable success has been made in the
application of the AdS/CFT duality to the physics of quark-gluon
plasma (QGP) created by RHIC, even though the underlying dynamics of
QCD is very different from that of a supersymmetric Yang-Mills
theory \cite{DTSon1, HLiu1, Yaffe}. It is the object of this brief
report to calculate $T_d$ using the heavy quark potential of $N=4$
SUSY YM extracted from its gravity dual \cite{maldacena2,SJRey,
HLiu2}.

Although the potential model applies only in the non-relativistic
limit which is not the case when the 't Hooft coupling, $\lambda$,
becomes too strong, it can be justified within the lower side of the
range of $\lambda$ used in the literature to compare AdS/CFT with
the RHIC phenomenology, i.e.$ 5.5<\lambda<6\pi.$  The upper edge is
obtained by using $N_c=3$ and the QCD value of $g_{\rm YM}$ at RHIC
energy scale ($g_{\rm YM}^2/(4\pi)\simeq 1/2$) and the lower edge is
based on a comparison between the heavy quark potential from lattice
simulation with that from AdS/CFT \cite{gubser}.

We model the quarkonium, $J/\psi$ or $\Upsilon$, as a
non-relativistic bound state of a heavy quark and its antiparticle .
The wave function for their relative motion satisfies the
Schr\"odinger equation
\begin{equation}
-\frac{1}{2m}\nabla^2\psi+V_{\rm eff.}(r)\psi = -E\psi
\label{schroedinger}
\end{equation}
where $m=M/2$ is the reduced mass with $M$ the mass of the heavy quark and $E(\ge 0)$ is the
binding energy of the bound state. Because of the
screening of QGP, the effective potential energy has a finite range and is temperature
dependent. The dissociation temperature of a particular state is the temperature at
which its energy, $-E$, is elevated to zero.

The free energy of a static pair of $q\bar q$ separated by a
distance $r$ at temperature $T$ is given
\begin{equation}
e^{-\frac{1}{T}F(r,T)}=\frac{{\rm tr}<W^\dagger(L_+) W(L_-)>}{{\rm
tr}{<W^\dagger(L_+)>< W(L_-)>}} \label{define}
\end{equation}
where $L_{\pm}$ stands for the Wilson line running in Euclidean time
direction at spatial coordinates $(0,0,\pm\frac{1}{2} r)$ and is
closed by the periodicity. We have  $ W(L_\pm)\equiv
Pe^{-i\int_0^{\frac{1}{T}}dtA_0(t,0,0,\pm\frac{1}{2}r)}$  with $A_0$
the temporal component of the gauge potential subject to the
periodic boundary condition, $A_0(t+\frac{1}{T},\vec r)=A_0(t,\vec
r)$. The trace here is over the color indexes and $<...>$ denotes
the thermal average. The symbol $P$ enforces the path ordering. The
corresponding internal energy reads $ U(r,T) =
=-T^2\frac{d}{dT}\left(\frac{F(r,T)}{T}\right)$.  Two ansatz of the
potential model have been explored in the literature\cite{karsch}:
the $F$-ansatz which identifies $V_{\rm eff.}$ of
(\ref{schroedinger}) with $F(r,T)$ and the $U$-ansatz which
identifies $V_{\rm eff.}$ with $U(r,T)$. The lattice QCD simulation
reveals that the $U$ ansatz produces a deeper potential well and
thereby higher $T_d$ because the entropy contribution is subtracted.
This remains the case with holographic potential as we shall see.

According to the holographic principle, the thermal average of a
Wilson loop operator $W(C)=Pe^{-i\oint_C dx^\mu A_\mu(x)}$  in 4D
$N=4$ SUSY YM at large $N_c$ and large 't Hooft coupling corresponds
to the minimum area $S_{\rm min.}[C]$ of the string world sheet in
the 5D AdS-Schwarzschild metric with a Euclidean signature ,
\begin{equation}
ds^2=\pi^2T^2y^2(fdt^2+d\vec x^2)+\frac{1}{y^2f}dy^2,
\label{metric}
\end{equation}
bounded by the loop $C$ at the boundary, $y\to\infty$, where
$f=1-\frac{1}{y^4}$. Specifically, we have
\begin{equation}
{\rm tr}<W(C)>=e^{-\sqrt{\lambda}S_{\rm min}[C]}.
\label{duality}
\end{equation}

For the numerator of (\ref{define}), $C$ consists of two parallel
temporal lines $(t,0,0,\pm\frac{r}{2})$ and the string world sheet
can be parameterized by $t$ and $y$ with the ansatz $x_1=x_2=0$ and
$x_3=$ a function of $y$. The induced world sheet metric reads
\begin{equation}
ds^2=\pi^2T^2y^2fdt^2+\Big[\pi^2T^2y^2\Big(\frac{dx_3}{dy}\Big)^2+\frac
{1}{\pi^2 T^2 y^2 f}\Big]dy^2.
\end{equation}
Minimizing the world sheet area (the Nambu-Goto action)

\begin{equation}
S[C]=(\pi T)\int_0^{\frac{1}{T}} dt\int_0^\infty dy\sqrt{1+\pi^4T^4y^4 f
\Big(\frac{dx_3}{dy}\Big)^2}
\end{equation}
generates two types of solutions\cite{maldacena2} \cite{SJRey}
\cite{HLiu2}. One corresponds to a single world-sheet with a
nontrivial profile $x_3(y)$,
\begin{equation}
x_3=\pm \pi Tq\int_{y_c}^y\frac{dy^\prime}{\sqrt{(y^{\prime
4}-1)(y^{\prime 4}-y_c^4)}}
\end{equation}
where $q$ is a constant of integration determined by the boundary
condition
\begin{equation}
r=\frac{2 q}{\pi T}
\int_{y_c}^\infty\frac{dy}{\sqrt{(y^4-1)(y^4-y_c^4)}} \label{boundary1}
\end{equation}
with $y_c^4=1+q^2$. The corresponding value of $\sqrt{\lambda}
S[C]$ is denoted by $I_1$. The other solution consists of two
parallel world sheets with $x_3=\pm\frac{r}{2}$ extending to the
black hole horizon and the corresponding value of $\sqrt{\lambda}
S[C]$ is denoted by $I_2$. The latter solution corresponds to two
non-interacting static quarks in the medium and is equal to the
denominator of (\ref{define})

The free energy we are interested in reads
\begin{equation}
F(r,T)=T{\rm min}(I,0)
\label{FreeEnergy}
\end{equation}
where
\begin{equation}
I\equiv I_1-I_2=\sqrt{\lambda}\Big[
\int_{y_c}^\infty dy\left(\sqrt{\frac{y^4-1}{y^4-y_c^4}}-1\right)+1-y_c\Big].
\label{nambu1}
\end{equation}
Inverting eq.(\ref {boundary1}) to express $q$ in terms of $r$ and
substituting the result to (\ref{nambu1}), it was found that the
function $I$ consists of two branches, The upper branch is always
positive and is therefore unstable. The lower branch starts from
being negative for $r<r_0$ and becomes positive for $r>r_0$. Both
branches joins at $r=r_c>r_0$ beyond which the nontrivial solution
ceases to exist. Numerically, we have $r_0\simeq \frac{0.7541}{\pi
T}$ and $r_c\simeq \frac{0.85}{\pi T}$. Introducing a dimensionless
radial coordinate $ \rho=\pi T r$, we find that
\begin{equation}
F(r,T)=-\frac{\alpha}{r}\phi(\rho)\theta(\rho_0-\rho),
\end{equation}
where
$\alpha=\frac{4\pi^2}{\Gamma^4\left(\frac{1}{4}\right)}\sqrt{\lambda}
\simeq 0.2285\sqrt{\lambda}$, and $\phi(\rho)=-\rho I/(\pi\alpha)$
is the screening factor . We have $\phi(0)=1$ and $\phi(\rho_0)=0$
with $ \rho_0 = 0.7541$.

The small $\rho$ expansion of $\phi(\rho)$ is given by
\begin{equation}
\phi(\rho)=1-\frac{\Gamma^4\left(\frac{1}{4}\right)}{4\pi^3}\rho+\frac{3\Gamma^8\left(\frac{1}{4}\right)}
{640\pi^6}\rho^4+O(\rho^8). \label{expansion}
\end{equation}
On writing the wave function $\psi(\vec r)=u_l(\rho)Y_{lm}(\hat
r)$, the radial Schr\"odinger equation for a zero energy bound
state reads
\begin{equation}
\frac{d^2u_l}{d\rho^2}+\frac{2}{\rho}\frac{du_l}{d\rho}-\Big[\frac{l(l+1)}{\rho^2}+{\cal V}\Big]u_l=0
\label{radialeq}
\end{equation}
with ${\cal V}=MV_{\rm eff.}/(\pi^2T^2)$. We have
\begin{equation}
{\cal V}=-\frac{\eta^2}{\rho_0\rho}\phi(\rho)\theta(\rho_0-\rho)
\end{equation}
for the F ansatz and
\begin{equation}
{\cal V}=-\frac {\eta^2}{\rho_0\rho}\Big[\phi(\rho)-\rho\Big(\frac{d\phi}{d\rho}\Big)\Big]
\theta(\rho_0-\rho)
\end{equation}
for the U ansatz, where $ \eta = \sqrt{\frac{\alpha\rho_0 M}{\pi
T}}$

Note that the potential of the U-ansatz jumps to zero from below at
$\rho=\rho_0$, since the derivative of $\phi(\rho)$ is nonzero
there. For both ansatz, and the case with an infrared cutoff
discussed below, the solution to (\ref{radialeq}) is given by
\begin{equation}
u_l={\rm const.}\rho^{-l-1}
\end{equation}
at $\rho>\rho_0$ and by
\begin{equation}
u_l={\rm const.}\rho^l
\end{equation}
near the origin. The threshold value of $\eta$ at the dissociation
temperature, $\eta_d$, is determined by the matching condition at
$\rho=\rho_0$,
\begin{equation}
\frac{d}{d\rho}(\rho^{l+1}u_l)\mid_{\rho=\rho_0^-}=0.
\label{matching}
\end{equation}
It follows from $\eta$ that the dissociation temperature is given by
\begin{equation}
T_d=\frac{\alpha\rho_0M}{\pi\eta_d^2}=
\frac{4\pi\rho_0}{\Gamma^4\left(\frac{1}{4}\right)\eta_d^2}\sqrt{\lambda}M.
\label{melting}
\end{equation}

It is interesting to observe that the extrapolation of the first two
terms of (\ref{expansion}) vanishes at
$\rho=\rho_0^\prime=4\pi^3/\Gamma^4\left(\frac{1}{4}\right)
 \simeq 0.7178$,
 which is very close to the exact zero point, and
the third term of (\ref{expansion}) remains small there. This
suggests that the screening factor $\phi(\rho)$ can be well
approximated by a linear function
\begin{equation}
\bar\phi(\rho)\simeq 1-\frac{\rho}{\bar\rho_0}
\end{equation}
with $\bar\rho_0=\frac{1}{2}(\rho_0+\rho_0^\prime)\simeq 0.7359$.
The effective potential $V_{\rm eff.}$ is then approximated by a
truncated Coulomb potential. We have
\begin{equation}
{\cal
V}=-\frac{\eta^2}{\rho_0\rho}\Big(1-\frac{\rho}{\rho_0}\Big)\theta(\rho_0-\rho)
\end{equation}
for the F-ansatz and
\begin{equation}
{\cal V}=-\frac{\eta^2}{\rho_0\rho}\theta(\rho_0-\rho)
\end{equation}
for the U-ansatz, where the over bar of $\rho_0$ has been
suppressed.


The radial wave function of the F-ansatz under the truncated Coulomb approximation
can be expressed in terms of the confluent hypergeometric function of the 1st
kind for $\rho<\rho_0$, i.e.
\begin{equation}
u_l=\rho^l{}_1F_1(l+1-\frac{\eta}{2};2l+2;2\eta\frac{\rho}{\rho_0}).
\end{equation}
The matching condition (\ref{matching}) yields the secular equation for $\eta$,
\begin{equation}
2l+1-\eta+\eta\left(1-\frac{\eta}{2l+2}\right)
\frac{{}_1F_1(l+2-\frac{\eta}{2};2l+3;2\eta)}{{}_1F_1(l+1-\frac{\eta}{2};2l+2;2\eta)}
=0
\end{equation}
As $\eta$ is reduced from above, we expect the bound states of the
same $l$ to melt successively. Therefore the first positive root
corresponds to the minimum binding strength for a bound state of
angular momentum $l$ and the 2nd one to the threshold of the first
radial excitation. Knowing the values of these $\eta$'s, the
disassociation temperature can be computed from the formula
(\ref{melting}). For example, the threshold $\eta$ of the 1$S$
state, $\eta_{1S}\simeq 1.76$, which implies that
\begin{equation}
T_d\simeq 0.0173\sqrt{\lambda}M.
\label{scaleF}
\end{equation}

In case of the $U$-ansatz under the same approximation, we find that
\begin{equation}
u_l=\frac{1}{\sqrt{\rho}}J_{2l+1}\Big(2\eta\sqrt{\frac{\rho}{\rho_0}}\Big)
\end{equation}
for $\rho<\rho_0$ with $J_\nu(x)$ the Bessel function. The secular equation for $\eta$ reads
\begin{equation}
2l+1-\eta\frac{J_{2l+2}(2\eta)}{J_{2l+1}(2\eta)}=0.
\end{equation}
We have $\eta_{1S}=1.20$ and
\begin{equation}
T_d\simeq 0.0370\sqrt{\lambda}M.
\label{scaleU}
\end{equation}

Numerical results for the dissociation temperature of quarkonium are
tabulated in table 1, where we have used the mass values
$M=1.65$GeV, $4.85$ GeV for $c$ and $b$ quarks .
The  errors caused by the truncated Coulomb approximation are
within 4 percent, as is shown by the numerical solution to the
Schr\"odinger equation of the exact potential.

\begin{table}
\begin{tabular}{|c|r|r|r|r|r|r|}
\hline
 ansatz\kern8pt& $J/\psi(1S)$
\kern8pt& $J/\psi(2S)$ \kern8pt& $J/\psi(1P)$
\kern8pt& $\Upsilon(1S)$ \kern8pt& $\Upsilon(2S)$ \kern8pt& $\Upsilon(1P)$ \kern8pt\\
\hline
$F$ \kern8pt&67-124 \kern8pt& 15-28 \kern8pt& 13-25
\kern8pt& 197-364 \kern8pt& 44-81 \kern8pt& 40-73 \kern8pt\\
\hline
$U$ \kern8pt& 143-265 \kern8pt& 27-50 \kern8pt& 31-58
\kern8pt& 421-780 \kern8pt& 80-148 \kern8pt& 92-171 \kern8pt\\
\hline
\end{tabular}
\bigskip
\caption{$T_d$ in MeV's under the truncated Coulomb approximation.
The lower value of each entry corresponds to $\lambda=5.5$ and the
upper one to $\lambda=6\pi$.}
 \label{table-IIIa}
\end{table}


Because of the conformal invariance at quantum level, there is no
color confinement in $N=4$ SUSY YM even at zero temperature. In
order to simulate the confined phase of QCD at low temperature, an
infrared cutoff has to be introduced that suppress the contribution
of the $AdS$ horizon. Two scenarios explored in the literature are
the hard-wall model and the soft-wall models \cite{DTSon2,DTSon3}.
The gravity dual of the de-confinement transition is modeled as the
Hawking-Page transition from a metric without a black hole at
$T<T_c$ to that with a black hole at $T>T_c$.  Heavy quark potential
and the meson dissociation temperatures calculated with the
hard-wall model are identical to what calculated above with the
vanila AdS-Schwarzschild metric.

In case of the simplest soft-wall model (\cite{DTSon3}), a dilaton
is introduced and  the gravity dual of the free energy  is the
 given by

\begin{equation}
F=-T\frac{1}{16\pi G_5}\int d^4x\int_{\rho_0}^\infty dr
e^{-\frac{c}{\rho^2}}\sqrt{g}(R-12), \label{soft1}
\end{equation}
where $c$ is determined by the $\rho$-mass and the transition
temperature is predicted as $T_c\simeq 0.2459m_\rho$ \cite{herzog}.
A variant of the soft-wall scenario proposed in
ref.\cite{Oleg,JTYee},  admits a string frame metric with a
conformal factor, i. e.
\begin{equation}
ds^2=\frac{e^{bz^2}}{z^2}(fdt^2+d\vec x^2+f^{-1}dz^2),
\label{metric_yee}
\end{equation}
The value of $b=0.184{\rm GeV}^2$ was obtained by fitting the
lattice simulated transition temperature $T_c=186$MeV \cite{JTYee}.
Following the steps from (\ref{metric}) to (\ref{nambu1}),we can
calculate the effective potentials at both ansatz\cite{houren}.
 To determine the dissociation temperature in this case, we have to
solve the Schr\"odinger equation (\ref{radialeq}) numerically with
the numerically calculated heavy quark potential for both ansatz,
since the truncated Coulomb potential no longer approximates well.
The modified dissociation temperatures are tabulated in the table 2,
which show an significant increment in the vicinity of $T_c$. The
comparison between the ratios $T_d/T_c$ we calculated here with that
obtained from the lattice QCD is shown in table 3\cite{houren}.

\begin{table}
\begin{center}
\begin{tabular}{|c|r|r|r|}
\hline
ansatz\kern24pt& $J/\psi$ \kern24pt& $\Upsilon$ \kern24pt\\
\hline
$F$ \kern24pt& NA \kern24pt& 235-385 \kern24pt\\
\hline
$U$ \kern24pt& 219-322 \kern24pt& 459-780 \kern24pt\\
\hline
\end{tabular}
\bigskip
\caption{$T_d$ in MeV's for the 1$S$ state with the deformed metric.
"NA" means that there is no bound states above $T_c$ and the entry
for the $\Upsilon$ with $U$ ansatz and $\alpha=6\pi$ is taken from
the table I, since no significant increment is observed.}
\label{table-II}
\end{center}
\end{table}
\begin{table}
\begin{center}
\begin{tabular}{|c|r|r|r|r|r|}
\hline
 ansatz\kern8pt&$J/\psi$(holographic) \kern8pt& $J/\psi$(lattice)
\kern8pt& $\Upsilon$(holographic) \kern8pt& $\Upsilon$(lattice) \kern8pt\\
\hline
$F$ \kern8pt& NA \kern8pt& 1.1 \kern8pt& 1.3-2.1 \kern8pt& 2.3 \kern8pt\\
\hline
$U$ \kern8pt& 1.2-1.7 \kern8pt& 2.0 \kern8pt& 2.5-4.2 \kern8pt& 4.5 \kern8pt\\
\hline
\end{tabular}
\bigskip
\caption{The ratio $T_d/T_c$ for the 1S from the holographic
potential
and that from the lattice QCD} \label{table-III}
\end{center}
\end{table}

In summary, we have calculated the dissociation temperatures of
heavy quarkonia using the static potential implied by the
holographic principle with both the vanila AdS-Schwarzschild metric
and the one with an infrared cutoff. While estimations of $T_d$ have
been made in the literature based on various holographic models
\cite{JunLi,Kim, Hoyos}, a determination of $T_d$ from the
Schr\"odinger equation within the same framework remains lacking.
Our work is to fill this gap. The authors of \cite{JunLi} gave an
order of magnitude estimate of the dissociation temperature relying
on the screening length only. The author of \cite{Kim} generalized
the spectral analysis of the light mesons to heavy mesons. Their
criterion for the dissociation, however, appears slightly ad hoc and
is again independent of the coupling. Both the screening length and
the coupling strength ought to affect the heavy quaronium binding.
Carrying out the analysis of the potential model inspired by the
holographic principle to the same extent of that of QCD will address
both contributions, especially the consistency of the range of the
coupling constant extracted from the jet quenching with the heavy
quarkonium physics. Also a detailed bound state calculation enables
us to assess the validity of the non-relativistic approximation
behind the potential model. On comparing our results with that from
the lattice simulation \cite{karsch}, we found that our ratios
$T_d/T_c$ extracted from the modified AdS-Schwarzschild metric
(\ref{metric_yee}) are lower than the lattice ones within a factor
of two. That the increment in $T_d/T_c$ from the F-ansatz to the
U-ansatz is about a factor of two is similar to what reported in
\cite{karsch}. One has to bear in mind that the lattice results
reviewed in \cite{karsch} were extracted from a pure $SU(3)$ gauge
theory without a matter field. On the other hand the matter field
contents of $N=4$ SUSY YM are larger than that of QCD with light
quarks. It is possible that the additional screening effect of the
matter field in $N=4$ SUSY YM makes the heavy quarkonia more
vulnerable and thereby lowers the dissociation temperature. This is
consistent with the observation that the potential well becomes
wider in the metric with the IR cutoff introduced in
\cite{Oleg,JTYee} since some of degrees of freedom becomes massive.

\section*{Acknowledgments}
We would like to thank D.T. Son,   J. T. Liu, H.Satz and R.
Mawhinney for discussions. The work of D.F.H. is supported in part
by Educational Committee under grants NCET-05-0675 and project No.
IRT0624.. The work of D.F.H and H.C.R. is also supported in part by
NSFC under grant Nos. 10575043,10735040 and by US Department of
Energy under grant DE-FG02-01ER40651-TaskB.

\end{document}